# Study Report of the Booster Transition Jump System

Xi Yang, Charles M. Ankenbrandt, William A. Pellico, and James Lackey
*Fermi National Accelerator Laboratory*
Box 500, Batavia IL 60510

## Abstract

There are more than 15 years since the transition-jump system (TJS) was installed. Because of the quad steering, which is caused by the beam not being well centered through all the $\gamma_T$ quads, the TJS has never been used in the operation. Based upon the recent calculation, the short 12 $\gamma_T$ quad had the worst offset in the vertical direction, and it was lowered 4 mm. Afterwards, the orbit error caused by the $\gamma_T$ quad steering has been largely removed, and this encourages our efforts of commissioning the TJS to reach its design goal.

## Introduction

The TJS has been installed in the Fermilab Booster since 1987 for the purpose of reducing the deleterious effects of passing through transition at high intensity by reducing the time that the beam spends near the transition energy.[1,2] However, since those $\gamma_T$ quads are not well aligned relative to the usual closed orbit, quad steering can cause beam loss, especially for high intensity beams, and a dispersion wave after transition. This is the major reason why the TJS has never been used in the operation since its installation.

A program, which uses the difference in the closed orbits when $\gamma_T$ quads are on and off and calculates the offsets of the beam relative to $\gamma_T$ quads, has been developed and tested.[3] Besides, a radial orbit offset (ROF) has been experimentally applied in finding the optimal radial position for centering the beam through all the $\gamma_T$ quads, thereby eliminating the immediate need for repositioning the quads in the horizontal direction. Unfortunately, any vertical offset requires the reposition of the beam relative to the $\gamma_T$ quad either by applying a local three-bump to the beam or by moving the $\gamma_T$ quad.

## Experimental Results

Based upon the calculated vertical offsets,[3] as shown in Fig. 1(a), since the short 12 $\gamma_T$ quad was the worst one with an offset of 8 mm above the beam center, it was lowered 4 mm, which is the maximum amount allowed by the quad stand. Besides, the BPM (beam position monitor) used for controlling the ROF was recently moved from the section of long 18 to the section of long 20, the programmed ROF was again used to move the beam radially when the $\gamma_T$ quads were pulsed to establish the ROF setting where the beam was best centered through all the $\gamma_T$ quads. The optimal ROF setting has changed from 3 to -4.5. All the measurements were done at the extracted beam intensity of $0.75\times10^{12}$ protons. The difference orbit between $\gamma_T$ quads on and off was measured at six different ROF values, -5, -4, -3, -2, -1, 0; the horizontal and vertical difference orbits are shown in Figs. 2(a) and 2(b) separately. All the orbits with $\gamma_T$ quads on were taken at the same $\gamma_T$ quad current with those in Fig. 1(a), which is 780 A. The black, red, green, blue, cyan, magenta curves represent six different ROF values of -5, -4, -3, -2, -1, and 0 respectively. It is clear that the beam is best centered through all the $\gamma_T$ quads at the ROF value of -4.5 (the middle of -5 and -4). The calculated vertical offsets of the beam relative to $\gamma_T$ quads are shown in Fig. 2(c). Comparing the vertical offset before and after the short 12 $\gamma_T$ quad movement, as shown in Fig. 1(a) and Fig. 2(c), the offset at short 12 has changed from -8 mm to -4 mm, which is consistent with the amount of the movement.

Furthermore, one expects that better the beam is centered through all the $\gamma_T$ quads, less the difference orbit between $\gamma_T$ quads on and off depends upon the peak value of the $\gamma_T$ waveform, and smaller the difference orbit will be. The ROF was set to the optimal value of -4.5. The difference orbit was taken at six different $\gamma_T$ quad settings of 1.75 kV, 2.0 kV, 2.25 kV, 2.5 kV, 2.75 kV, and 3.0 kV. The horizontal and vertical difference orbits are shown in Fig. 3(a) and 3(b). The black, red, green, blue, cyan, magenta curves represent six different $\gamma_T$ quad settings of 1.75 kV, 2.0 kV, 2.25 kV, 2.5 kV, 2.75 kV, and 3.0 kV respectively. Since ideally the difference orbit between $\gamma_T$ quads on and off should be zero, the difference orbit can be represented as the orbit error either in average or in rms. The horizontal and vertical orbit errors are shown in Figs. 3(c) and 3(d). The black and red curves represent the orbit error in average and in rms respectively. There is a slight dependence between the horizontal orbit error in rms and the $\gamma_T$ quad setting, and the horizontal orbit error in rms is about 0.1 mm. Comparably, the vertical orbit error in rms is independent of the $\gamma_T$ quad setting, and is about 0.06 mm. So till now, the beam is well centered through all the $\gamma_T$ quads at the ROF setting of -4.5, which was used for all the following experiments.

There are three important parameters, which need to be adjusted in the experiment, for the purpose of minimizing the time that the beam spends near the transition energy, compensating the phase-space mismatch caused by space charge at transition, and minimizing the beam loss when the $\gamma_T$ quads are pulsed. These three parameters are the $\gamma_T$ quad setting ($V$), the trigger time ($T_P$) for pulsing $\gamma_T$ quads, and the transition gate ($T_G$). $V$ determines the amount of the reduction in $\gamma_T$ ($\Delta\gamma_T$). $\Delta\gamma_T$ decides how earlier $T_G$ could be relative to the normal transition gate ($t_g$). Earlier $T_P$ is, earlier $T_G$ is, and less time the beam spends near the transition energy. Since $T_G$ must be set to the time when the beam energy is equal or close to the transition energy, larger $\Delta\gamma_T$ is, lower the transition energy could be, and earlier $T_G$ could be. $T_G$ must be set in the range of $T_P$ to $T_P$+0.1 ms since it takes 0.1 ms, which is the rising time of the $\gamma_T$ waveform, for $\Delta\gamma_T$ to reach its maximum. Within this range, the delay between $T_G$ and $T_P$ is mainly determined by the optimal condition for compensating the phase-space mismatch caused by space charge. However, since the larger $\Delta\gamma_T$ is, the stronger the quad steering usually is, it is likely that a small portion of the high intensity beam will be lost when the $\gamma_T$ quads are pulsed. Since the optimal settings for parameters $V$, $T_P$, and $T_G$ are usually determined by their intrinsic relations and the beam intensity, they need to be optimized for different beam intensities via experiment.

According to the relations among $V$, $T_P$, and $T_G$, $V$ needs to be determined first, and it should be set to the maximum value before any beam loss starts. Afterwards, fixing the time delay between $T_G$ and $T_P$ to a reasonable value via experience, such as 20 μs, one adjusts $T_P$ to find the optimal setting based upon the criteria that there should be the least change in the bunch shape and bunch lengthen cross the transition. Finally, setting $T_P$ to its optimal value, one scans the delay between $T_G$ and $T_P$ in the range of 0 to 100 μs in a small step, such as 5 μs, to find the optimal setting for $T_G$ based upon the same criteria for the optimization of $T_P$. In the experiment, the bunch length and bunch shape are monitored by the resistive wall signal.

The above procedures are used to find the optimal settings of $V$, $T_P$, and $T_G$ for two different beam intensities.

First, the mountain range plot (MRP) was used to record the process of the transition jump at the extracted beam intensity of $0.75\times10^{12}$ protons. The MRP was triggered 100 μs before the transition, with 1 trace per 60 turns for 30 traces. $V$ could be set to a much higher value than those for high intensity beams, since in general, lower the beam intensity is, smaller the beam transverse size and emittance are, and higher the limit for the quad steering can be achieved. We took advantage of running the beam at a low intensity, experimentally investigated relations among $V$, $T_P$, and $T_G$, and found whether or not the TJS would help in reducing the beam load during the transition and gaining some effective accelerating voltages. From the MRP of Figs. 4(a) to 4(r), the optimal $T_G$ *vs*. $T_P$ at three different $V$ settings of 2.0 kV, as shown in Fig. 4(b), 2.5 kV, as shown in Fig. 4(i), and 3.0 kV, as shown in Fig. 4(q), are extracted using the criteria of the least bunch-length oscillation. The results are shown in Fig. 5(a), and are consistent with the above analysis, which indicates that $T_P$ should be set to an earlier time when $V$ is at a larger value. Besides, at a fixed $T_P$ of 18.4 ms, the optimal delays between $T_G$ and $T_P$ at two different $V$ of 2.5 kV, from Figs. 4(g) to 4(l), and 3.0 kV, from Figs. 4(s) to 4(u), are extracted, and shown in Fig. 5(b). Once the RF accelerating voltage curve is fixed, the synchronous phase determines the effective accelerating voltage,[4] and it was found to be independent of different settings for $V$, $T_P$, and $T_G$ in the experiment, as shown in Figs. 5(c) and 5(d).

The high intensity situation is quite different from the low intensity. The same procedures, which were used at the extracted beam intensity of $0.75\times10^{12}$ protons, were used for the extracted beam intensity of $4.0\times10^{12}$ protons. The maximum $V$ was found to be 1.5 kV. $T_P$ was fixed at its optimal value of 18.70 ms. $T_G$ was varied at eight different values of 18.71 ms, 18.72 ms, 18.725 ms, 18.73 ms, 18.735 ms, 18.74 ms, 18.75 ms, and 18.76 ms, and their MRP are shown in Figs. 6(a) to 6(h) respectively. The optimal $T_G$ was found to be somewhere between 18.725 ms (Fig. 6(c)) to 18.730 ms (Fig. 6(d)). MRP at the normal operational condition is shown in Fig. 6(i). Comparing Fig. 6(c) to Fig. 6(i), the bunch was shorter and the bunch length oscillation was much stronger during the transition in Fig. 6(i) than those in Fig. 6(c).

## Comment

The design goal of the TJS is to reduce the deleterious effects of high intensity beam passing through transition by reducing the time that the beam spends near the transition energy. It is experimentally demonstrated that the TJS can be used to reduce the peak current during the transition crossing and compensate the phase-space mismatch caused by the space charge via properly timing $T_P$ and $T_G$ to avoid the shortening and oscillation in the bunch length.

However, the emittance growth before transition, especially in the early part of the cycle when the space charge influence is the strongest, couldn't be cured by the TJS. Besides, from the present investigation, it is clear that the commission of the TJS couldn't reduce the requirement for the amount of the accelerating voltage at the transition; the beam intensity is mainly limited by the RF power unless the present RF system is upgraded. Furthermore, commissioning the TJS requires that the beam is well centered through all the $\gamma_T$ quads during the entire $\gamma_T$ waveform, which is about 4 to 5 ms. The higher the beam intensity is, the more strict this requirement becomes, and this puts a serious constrain to the orbit during the transition.

**Acknowledgments**

Thanks to Andrew Feld and Roy Mraz for helping the short 12 $\gamma_T$ quad movement. Also, thanks to Rich Meadowcroft and Andrew Feld for repairing the $\gamma_T$ power supply.

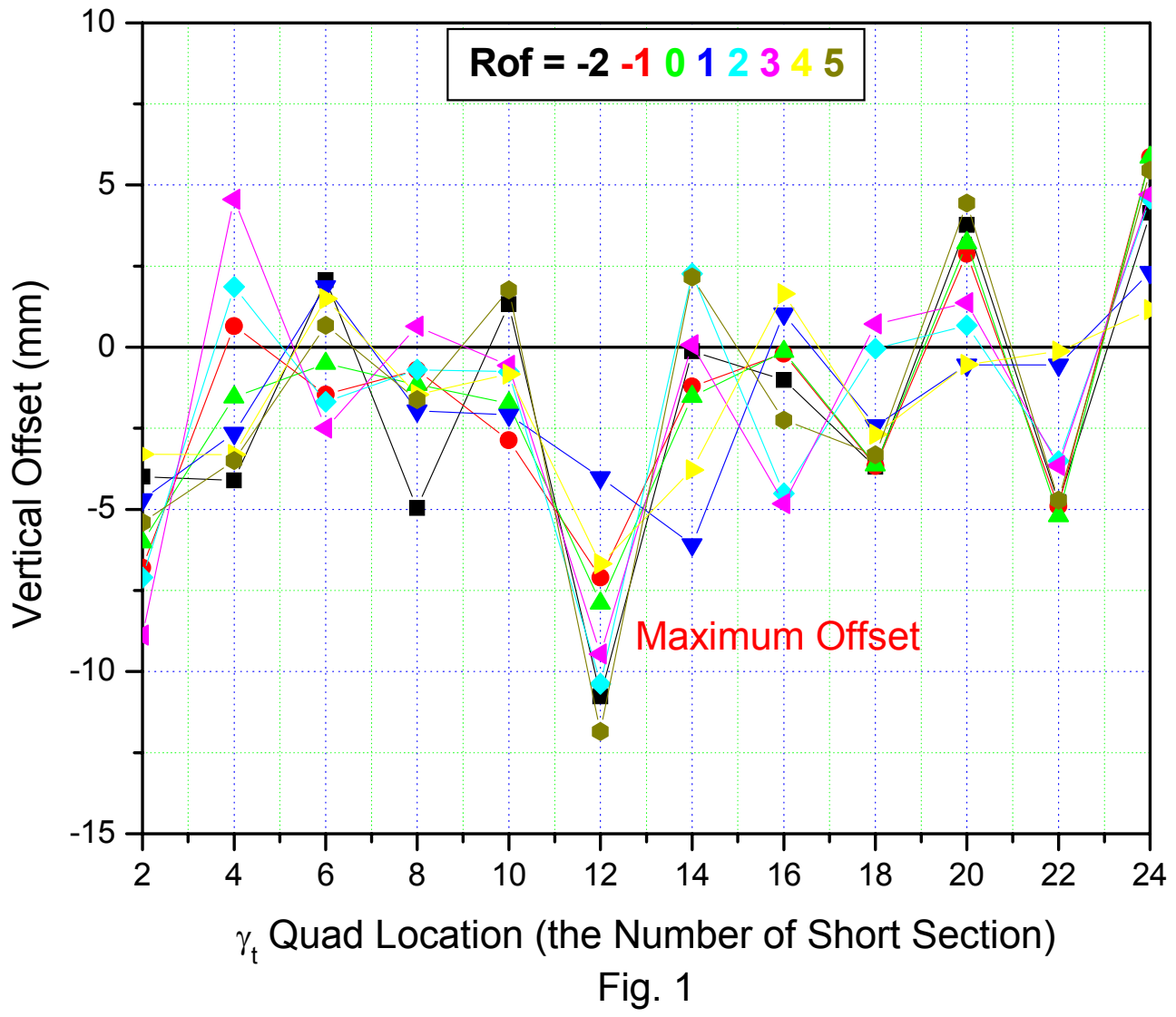

Fig. 1

Fig. 1 the calculated offsets of the beam relative to $\gamma_T$ quads in the vertical direction at ROF values of -2 to 5 before the movement of the short 12 $\gamma_T$ quad. The black, red, green, blue, cyan, magenta, yellow, dark yellow curves represent the eight different ROF values of -2, -1, 0, 1, 2, 3, 4, 5 respectively.

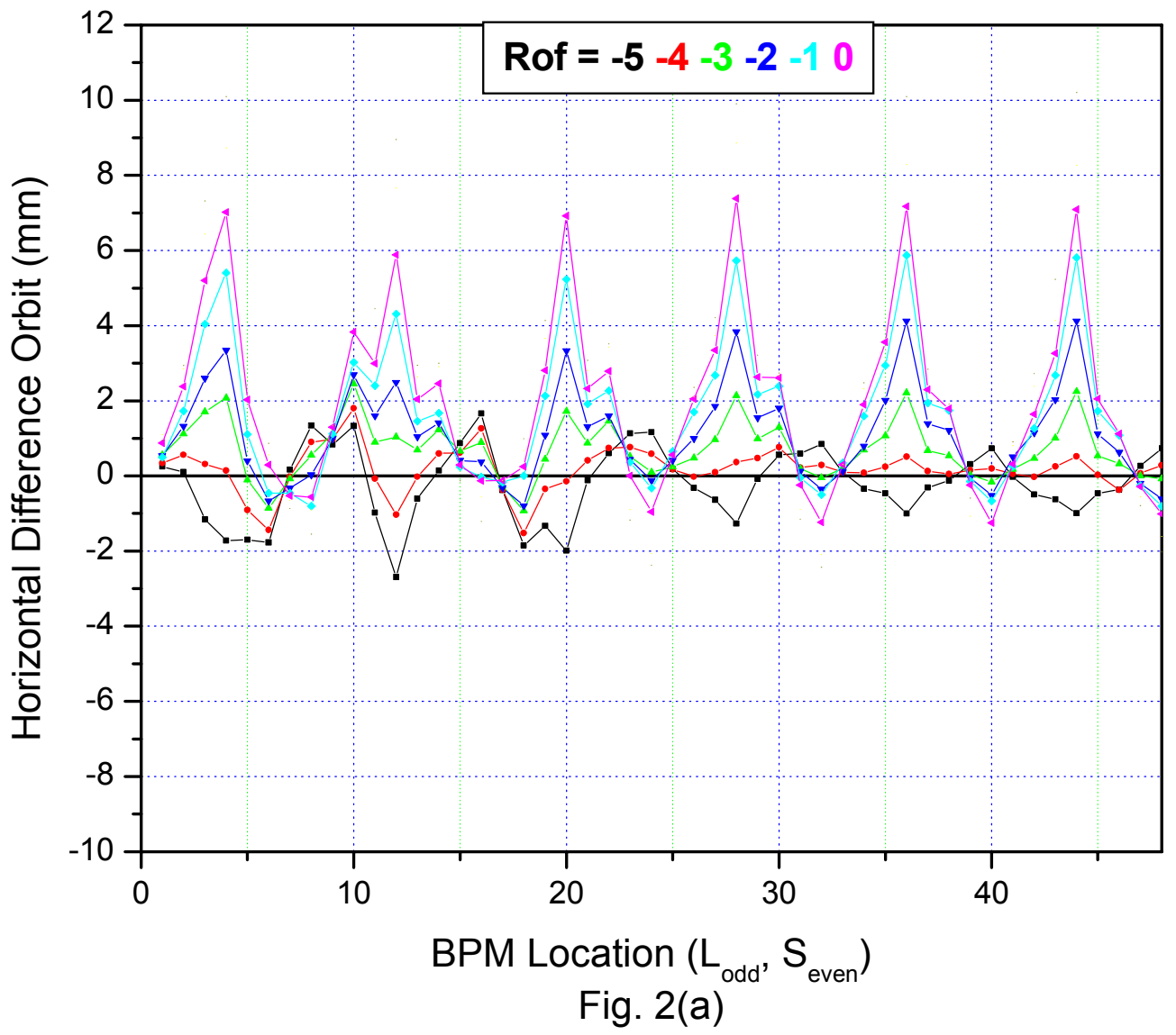

Rof = -5 -4 -3 -2 -1 0
Horizontal Difference Orbit (mm)
BPM Location ($L_{odd}$, $S_{even}$)

Fig. 2(a)

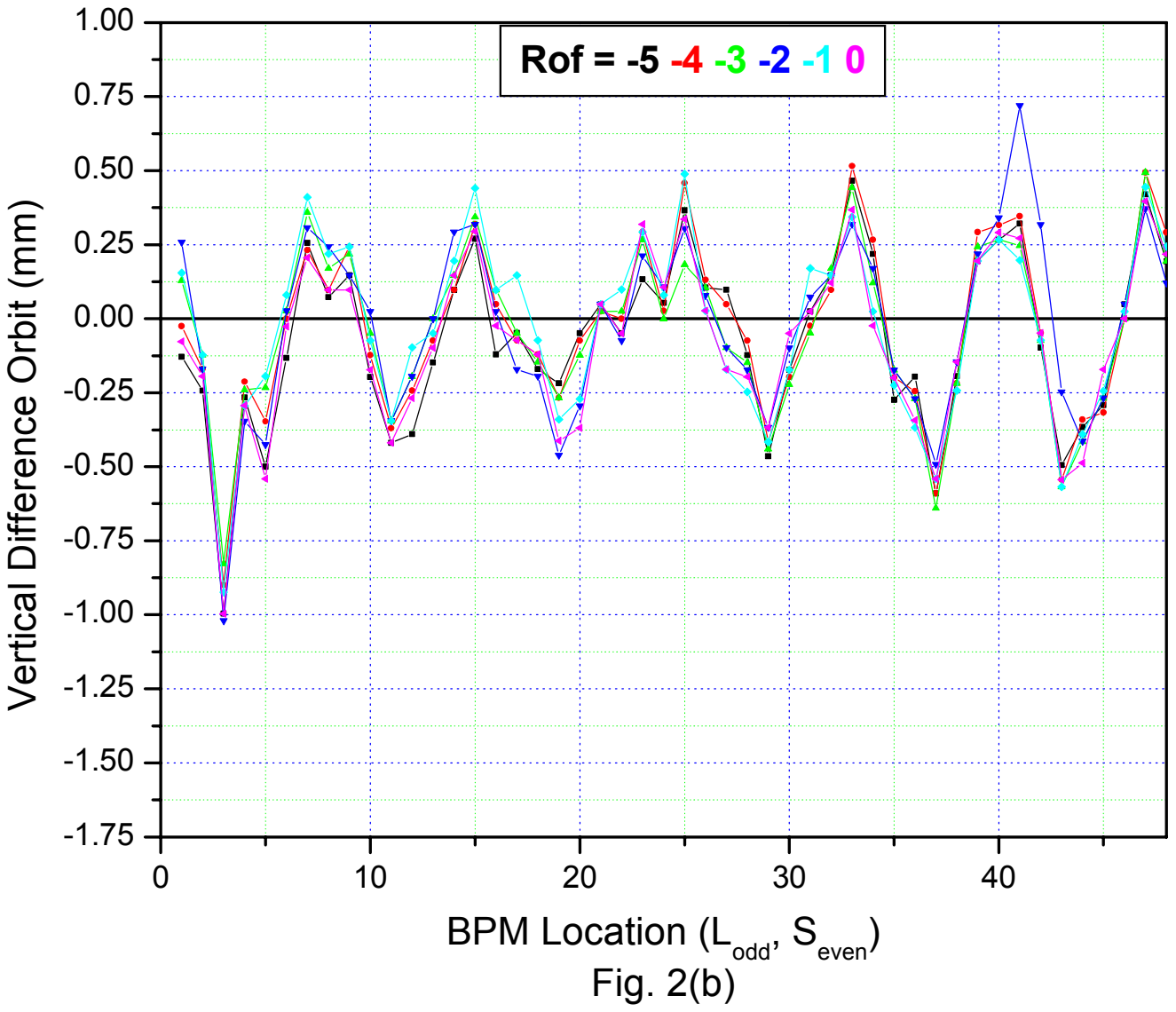

Rof = -5 -4 -3 -2 -1 0
Vertical Difference Orbit (mm)
BPM Location ($L_{odd}$, $S_{even}$)

Fig. 2(b)

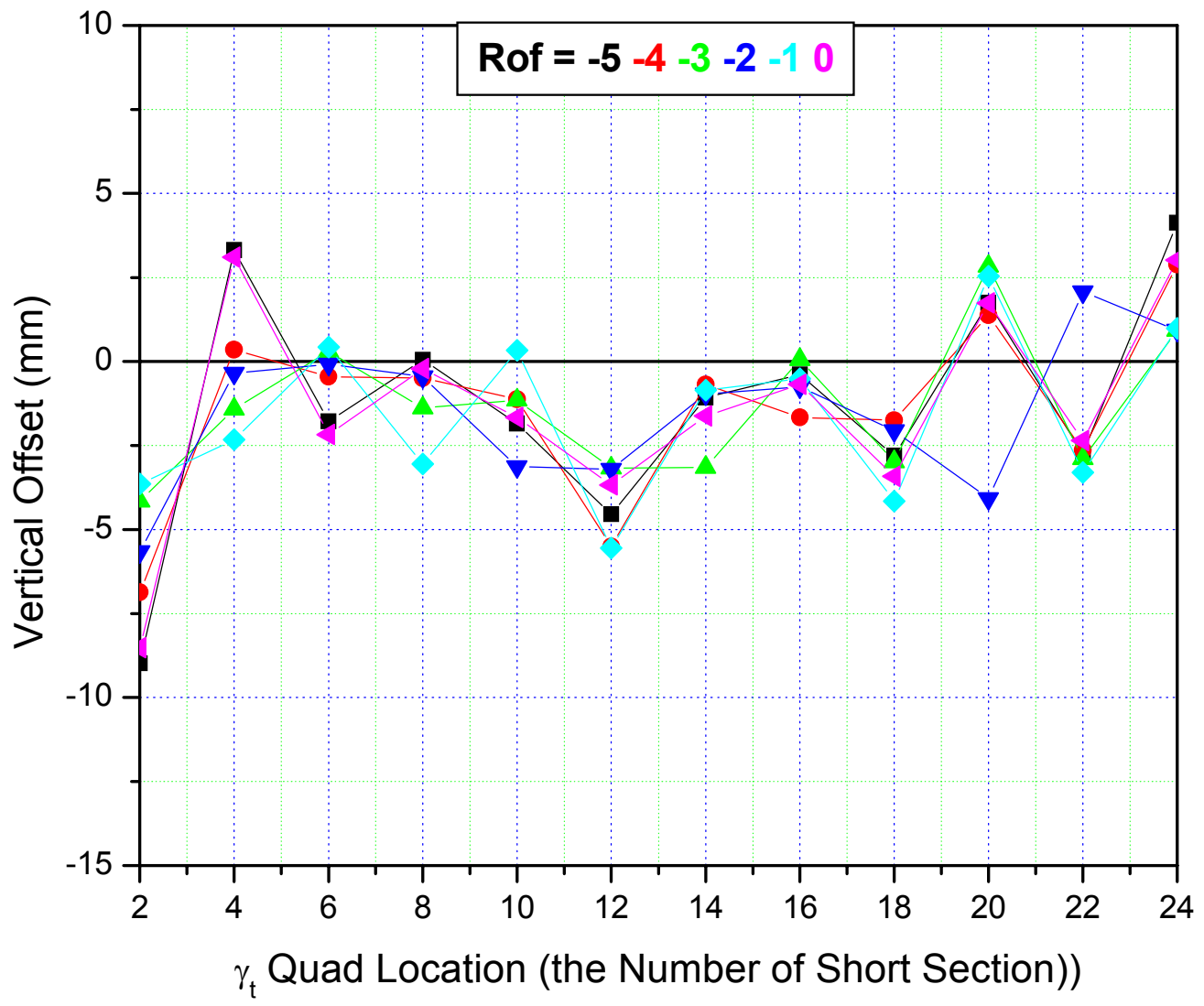

Fig. 2(c)

Fig. 2(a) after the short 12 $\gamma_T$ quad movement, the horizontal difference orbit with $\gamma_T$ quads on and off measured at six ROF values for the extracted beam intensity of $0.75\times10^{12}$ protons. The black, red, green, blue, cyan, magenta curves represent the six different ROF values of -5, -4, -3, -2, -1, and 0 respectively.

Fig. 2(b) the vertical difference orbit with $\gamma_T$ quads on and off measured at six ROF values for the extracted beam intensity of $0.75\times10^{12}$ protons.

Fig. 2(c) the calculated offsets of the beam relative to $\gamma_T$ quads in the vertical direction at ROF values of -5 to 0.

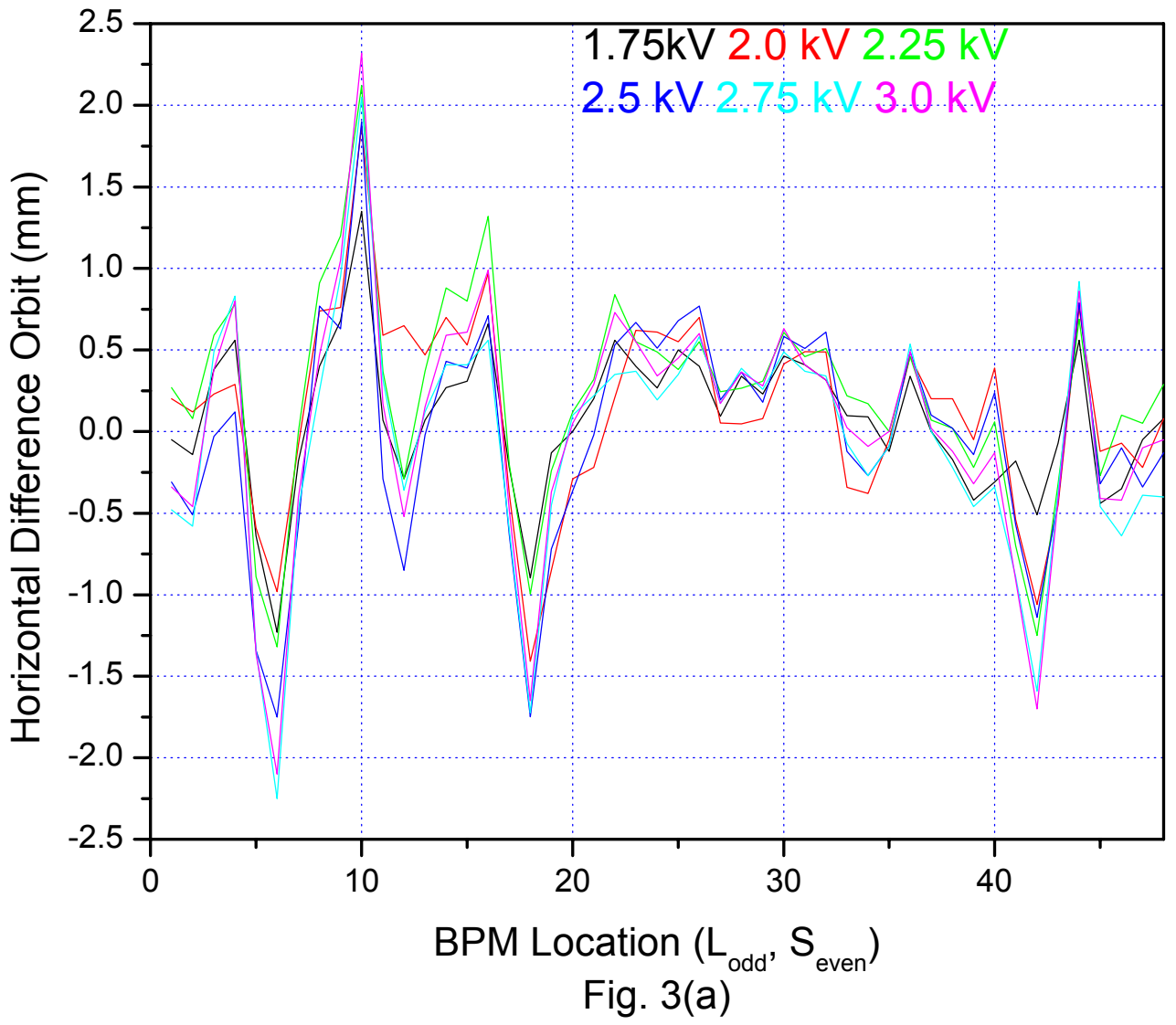

1.75kV 2.0 kV 2.25 kV
2.5 kV 2.75 kV 3.0 kV
Horizontal Difference Orbit (mm)
BPM Location ($L_{odd}$, $S_{even}$)

Fig. 3(a)

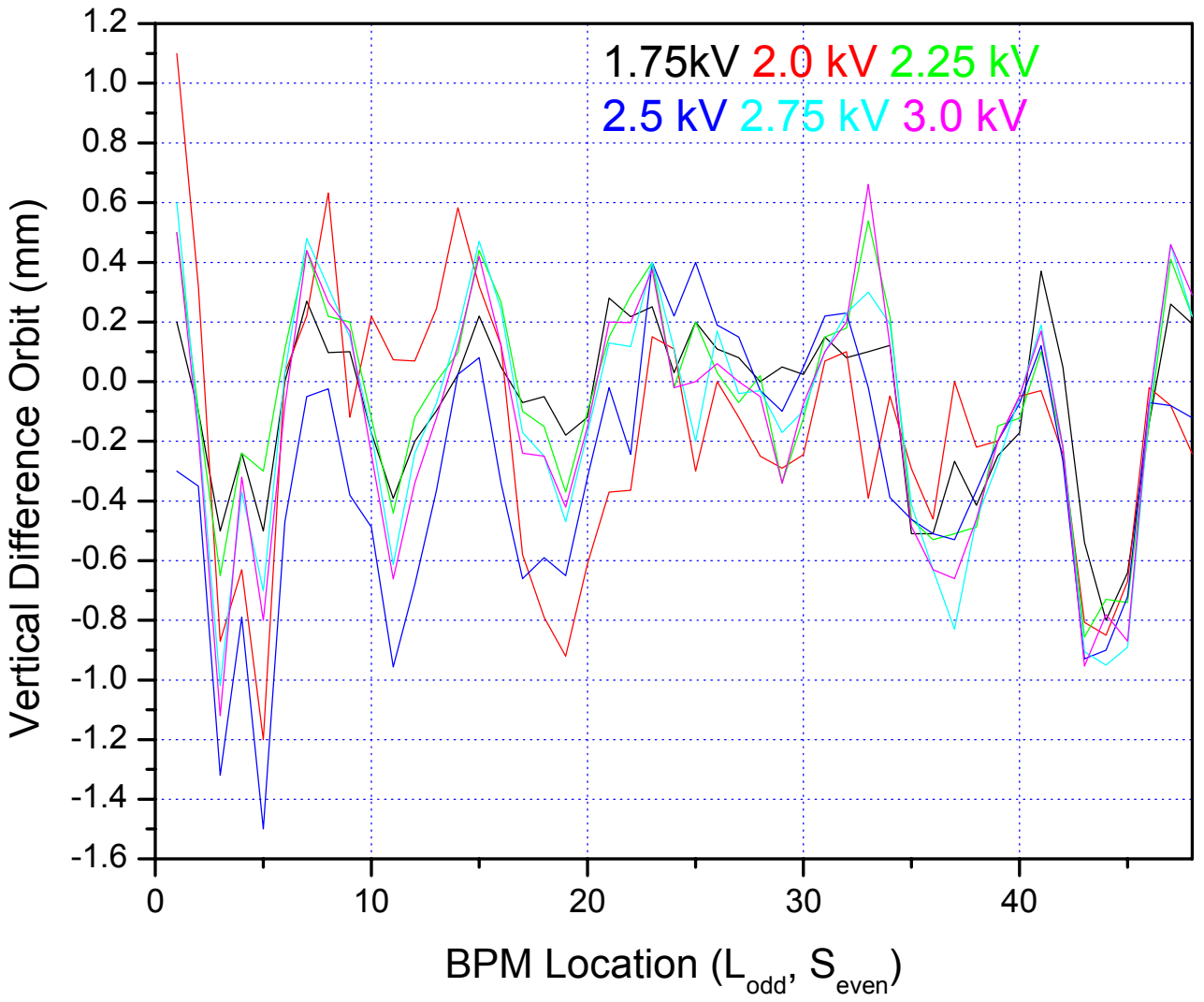

1.75kV 2.0 kV 2.25 kV
2.5 kV 2.75 kV 3.0 kV
Vertical Difference Orbit (mm)
BPM Location ($L_{odd}$, $S_{even}$)

Fig. 3(b)

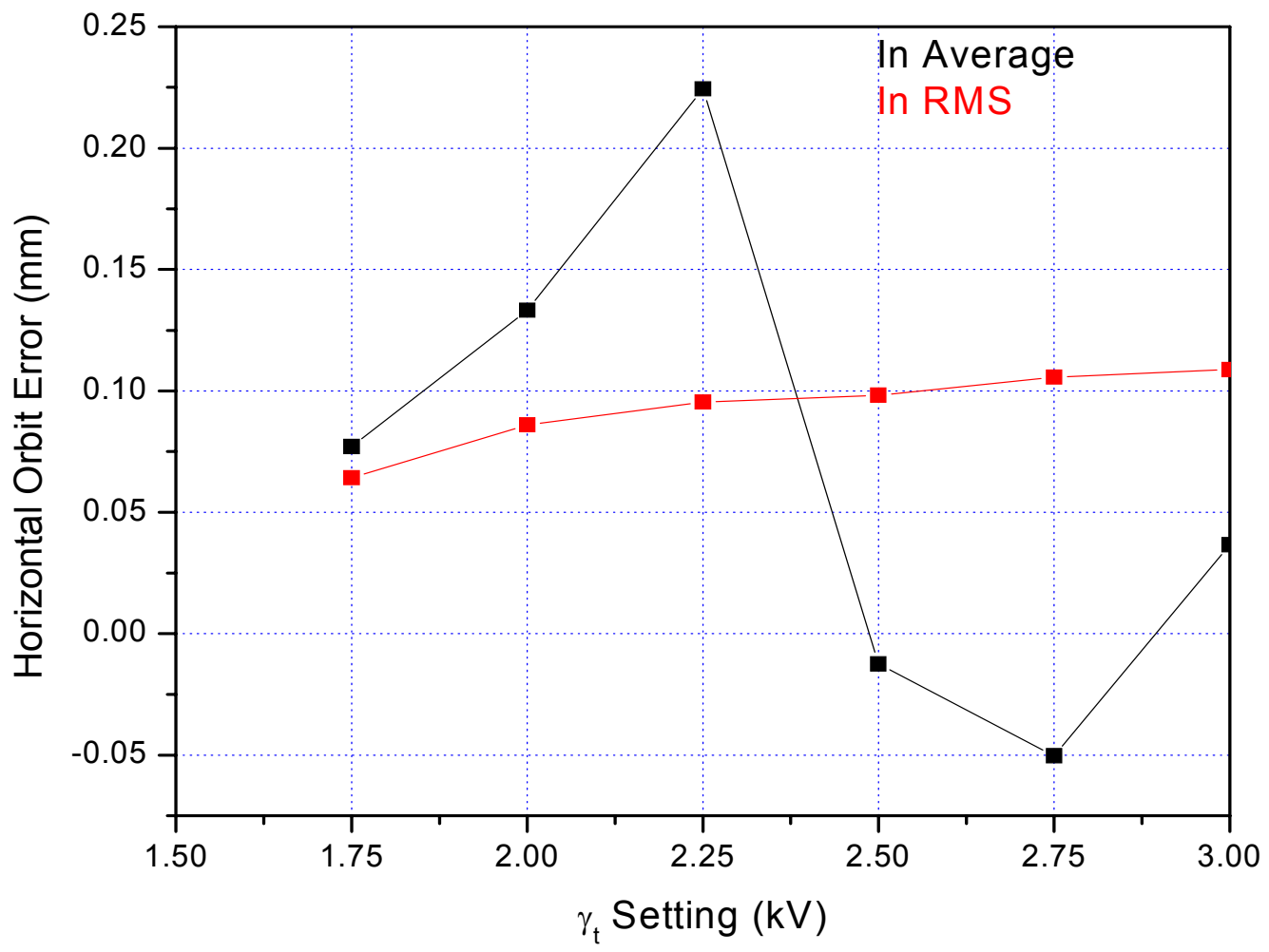

Fig. 3(c)

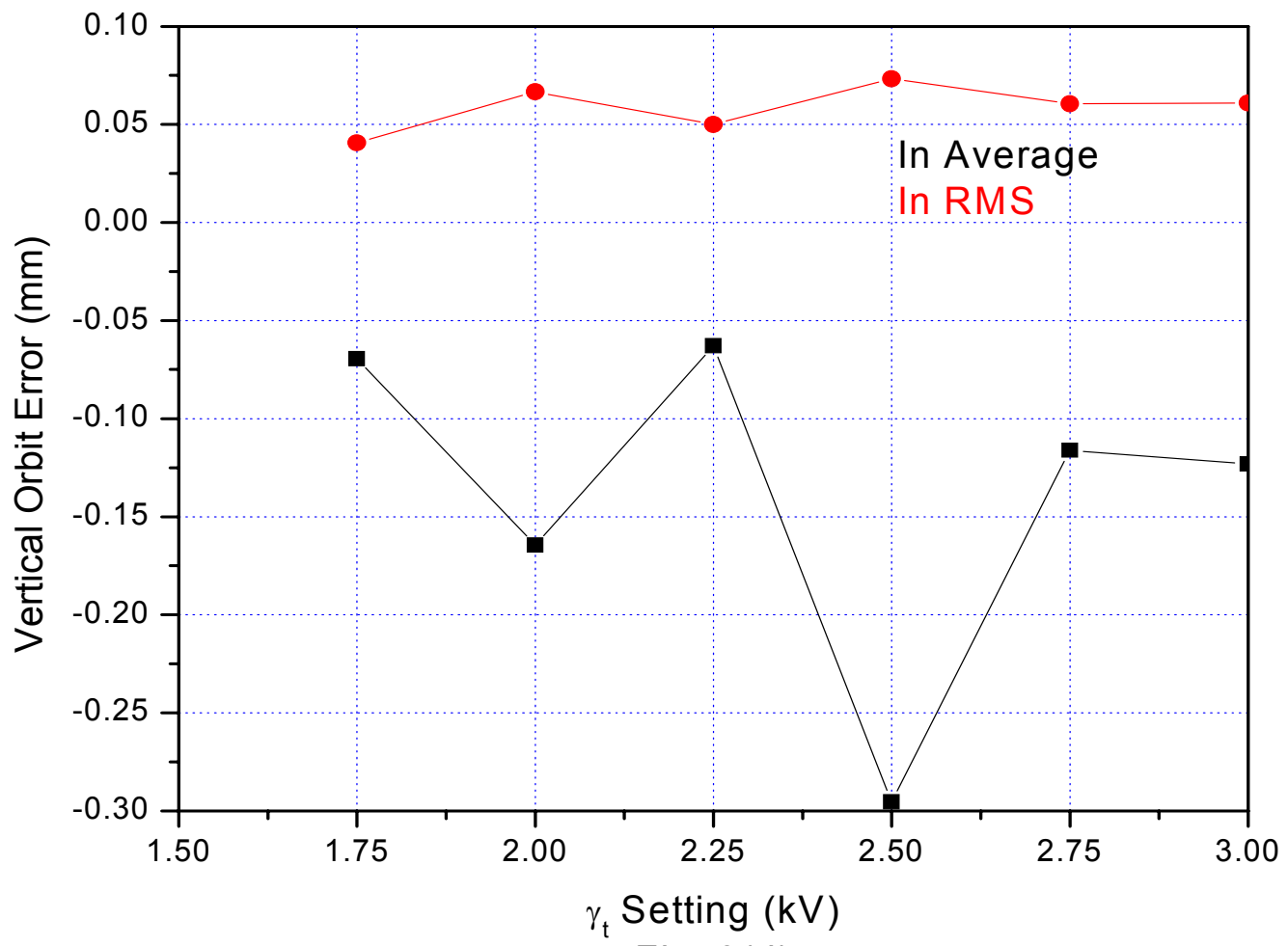

Fig. 3(d)

Fig. 3(a) at ROF=-4.5, the horizontal difference orbit with $\gamma_T$ quads on and off measured at six different $\gamma_T$ quad settings of 1.75 kV, 2.0 kV, 2.25 kV, 2.5 kV, 2.75 kV, and 3.0 kV. The black, red, green, blue, cyan, magenta curves represent the six different $\gamma_T$ quad settings of 1.75 kV, 2.0 kV, 2.25 kV, 2.5 kV, 2.75 kV, and 3.0 kV respectively.

Fig. 3(b) the corresponding vertical difference orbits with those in Fig. 3(a).

Fig. 3(c) the horizontal orbit errors, which correspond to the orbits in Fig. 3(a), in average (the black curve) and in rms (the red curve).

Fig. 3(d) the vertical orbit errors, which correspond to the orbits in Fig. 3(b), in average (the black curve) and in rms (the red curve).

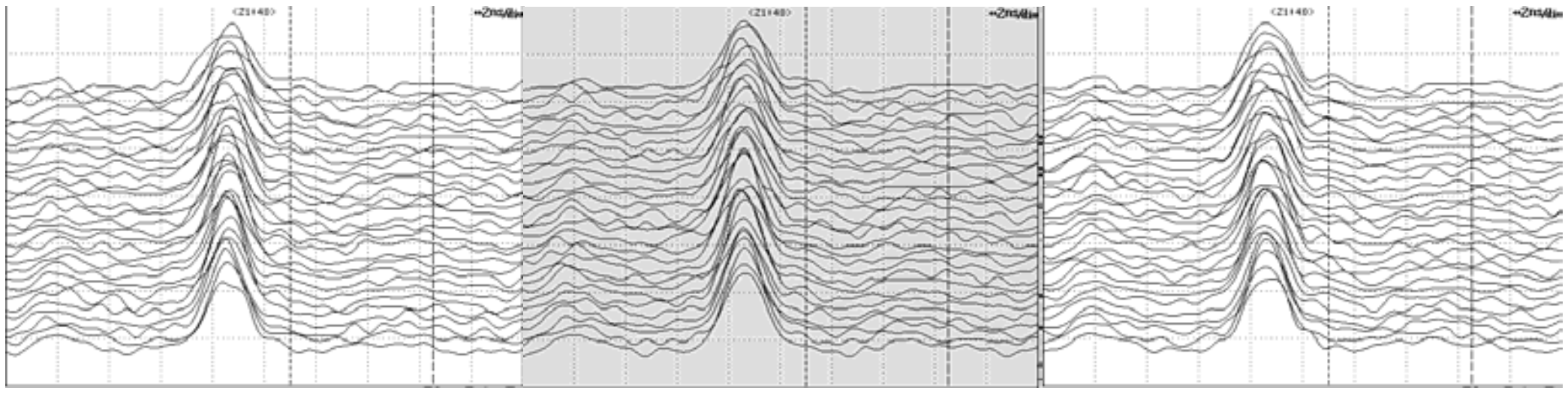

(a) (b) (c)

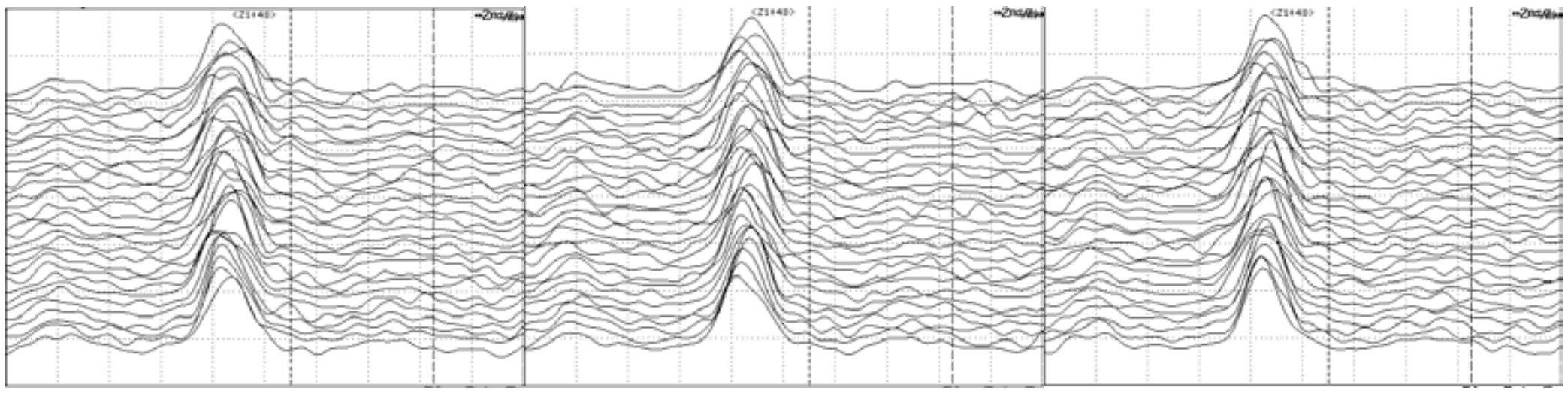

(d) (e) (f)

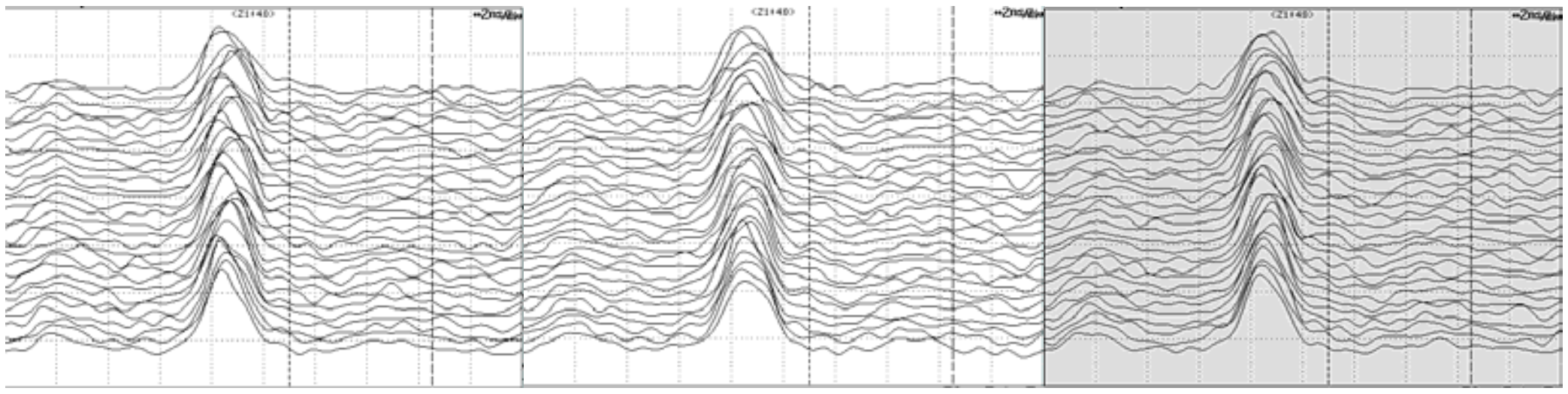

(g) (h) (i)

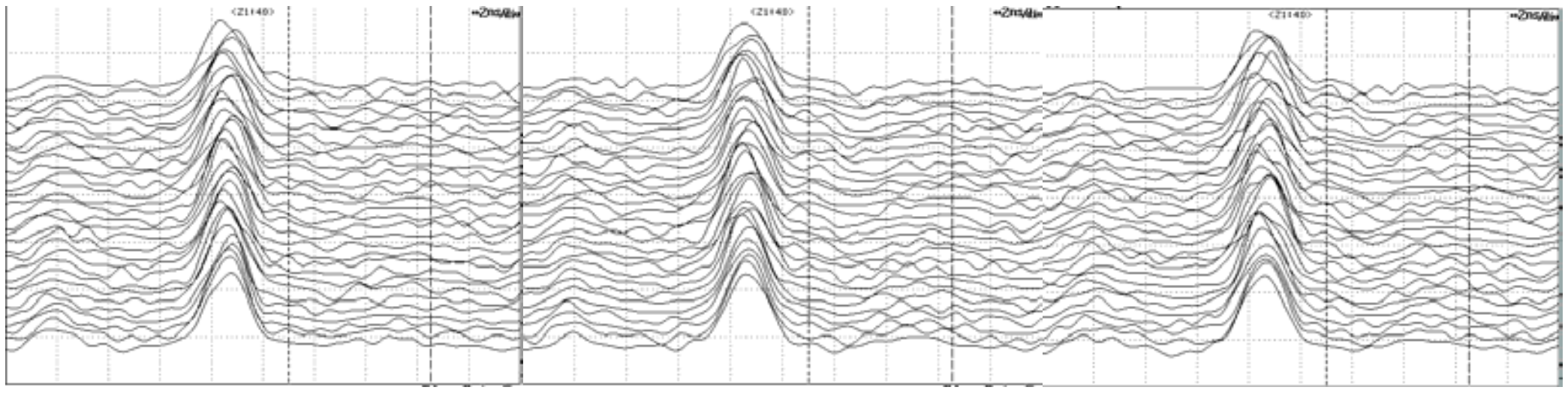

(j) (k) (l)

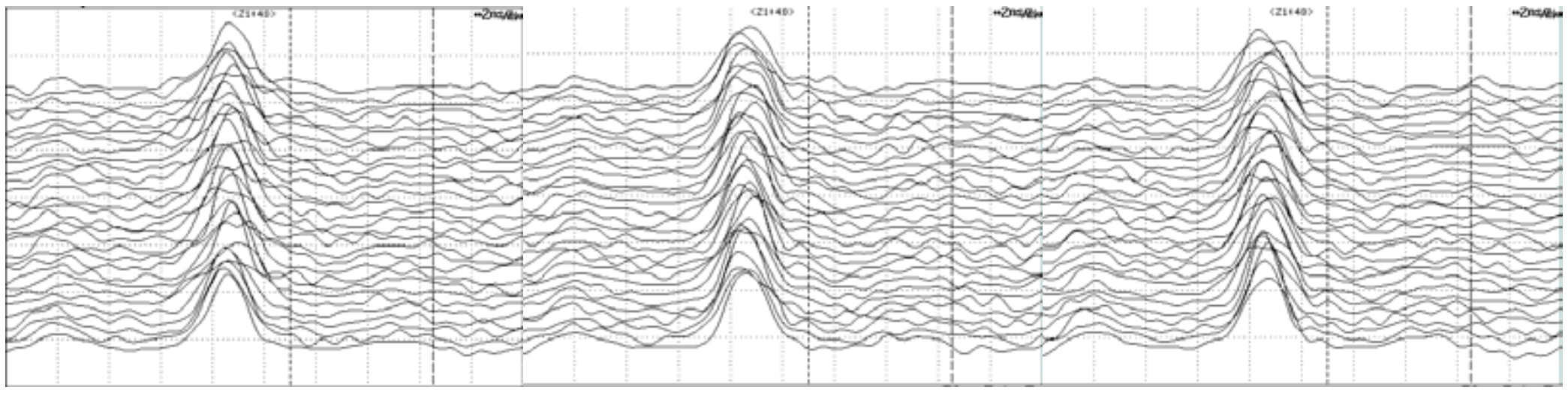

(m) (n) (o)

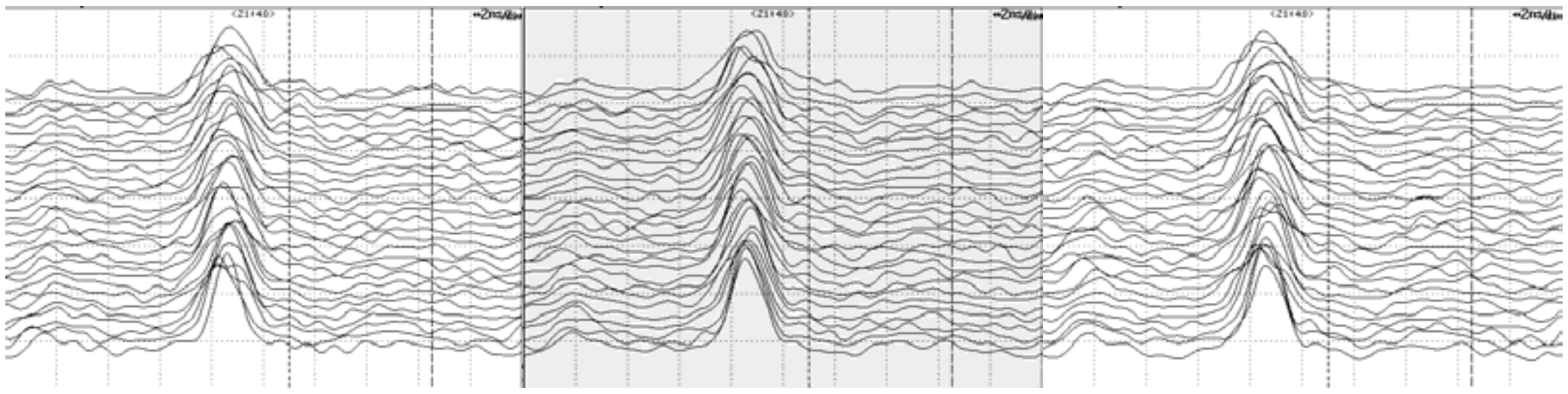

(p) (q) (r)

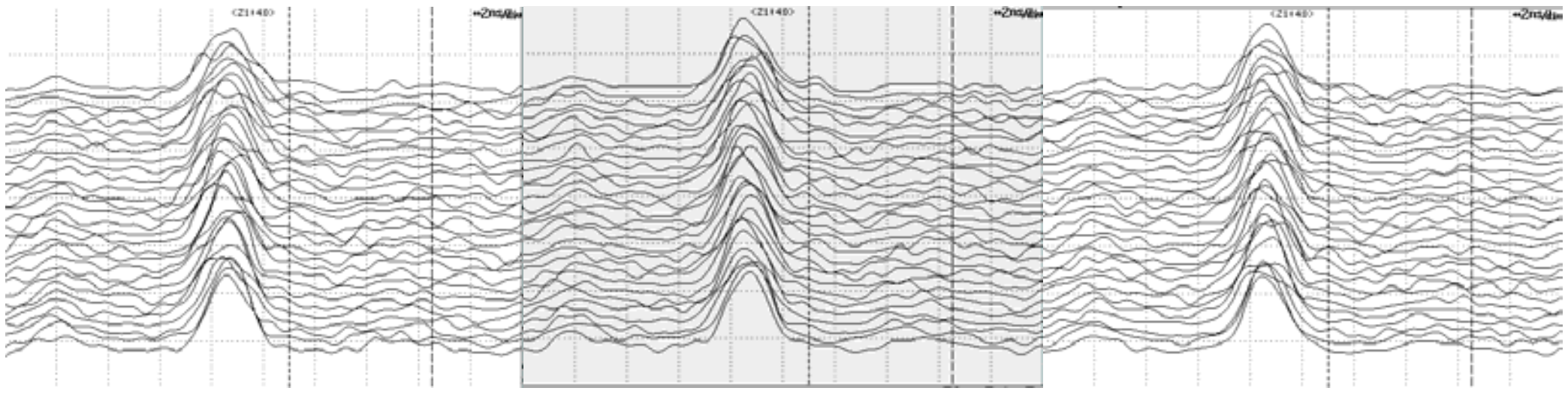

(s) (t) (u)

Fig. 4(a) at the extracted beam intensity of $0.75\times10^{12}$ protons, MRP took at the TJS setting of $V$=2.0 kV, $T_P$=18.5 ms and $T_G$=18.55 ms.

Fig. 4(b) MRP at the TJS setting of $V$=2.0 kV, $T_P$=18.55 ms and $T_G$=18.60 ms.

Fig. 4(c) MRP at the TJS setting of $V$=2.0 kV, $T_P$=18.60 ms and $T_G$=18.65 ms.

Fig. 4(d) MRP at the TJS setting of $V$=2.0 kV, $T_P$=18.65 ms and $T_G$=18.70 ms.

Fig. 4(e) MRP at the TJS setting of $V$=2.0 kV, $T_P$=18.70 ms and $T_G$=18.75 ms.

Fig. 4(f) MRP at the TJS setting of $V$=2.0 kV, $T_P$=18.75 ms and $T_G$=18.80 ms.

Fig. 4(g) MRP at the TJS setting of $V$=2.5 kV, $T_P$=18.40 ms and $T_G$=18.41 ms.

Fig. 4(h) MRP at the TJS setting of $V$=2.5 kV, $T_P$=18.40 ms and $T_G$=18.415 ms.

Fig. 4(i) MRP at the TJS setting of $V$=2.5 kV, $T_P$=18.40 ms and $T_G$=18.420 ms.

Fig. 4(j) MRP at the TJS setting of $V$=2.5 kV, $T_P$=18.40 ms and $T_G$=18.425 ms.

Fig. 4(k) MRP at the TJS setting of $V$=2.5 kV, $T_P$=18.40 ms and $T_G$=18.430 ms.

Fig. 4(l) MRP at the TJS setting of $V$=2.5 kV, $T_P$=18.40 ms and $T_G$=18.440 ms.

Fig. 4(m) MRP at the TJS setting of $V$=3.0 kV, $T_P$=18.60 ms and $T_G$=18.65 ms.

Fig. 4(n) MRP at the TJS setting of $V$=3.0 kV, $T_P$=18.30 ms and $T_G$=18.35 ms.

Fig. 4(o) MRP at the TJS setting of $V$=3.0 kV, $T_P$=18.10 ms and $T_G$=18.15 ms.

Fig. 4(p) MRP at the TJS setting of $V$=3.0 kV, $T_P$=18.05 ms and $T_G$=18.05 ms.

Fig. 4(q) MRP at the TJS setting of $V$=3.0 kV, $T_P$=18.05 ms and $T_G$=18.08 ms.

Fig. 4(r) MRP at the TJS setting of $V$=3.0 kV, $T_P$=18.05 ms and $T_G$=18.10 ms.

Fig. 4(s) MRP at the TJS setting of $V$=3.0 kV, $T_P$=18.40 ms and $T_G$=18.41 ms.

Fig. 4(s) MRP at the TJS setting of $V$=3.0 kV, $T_P$=18.40 ms and $T_G$=18.43 ms.

Fig. 4(s) MRP at the TJS setting of $V$=3.0 kV, $T_P$=18.40 ms and $T_G$=18.44 ms

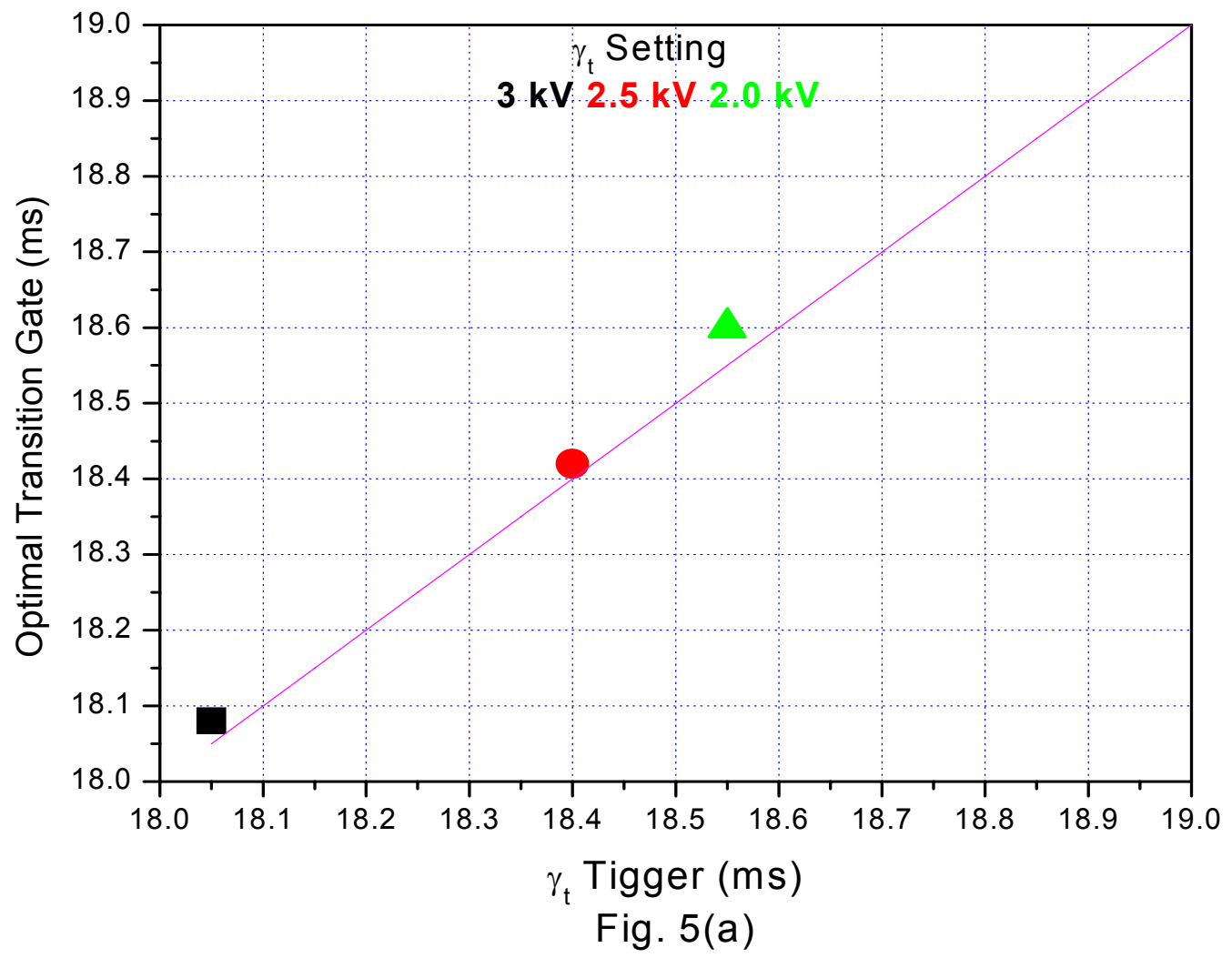

Fig. 5(a)

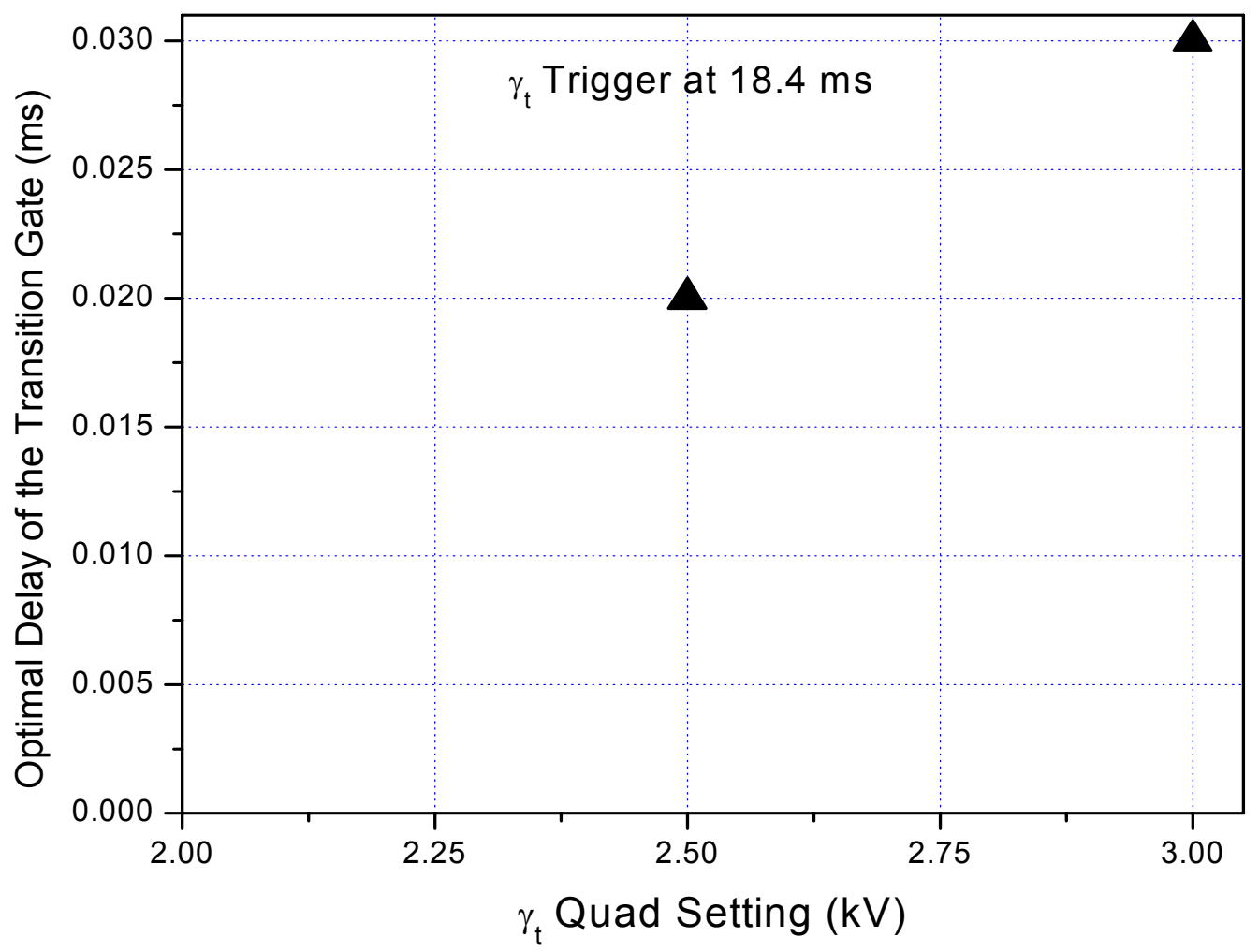

Fig. 5(b)

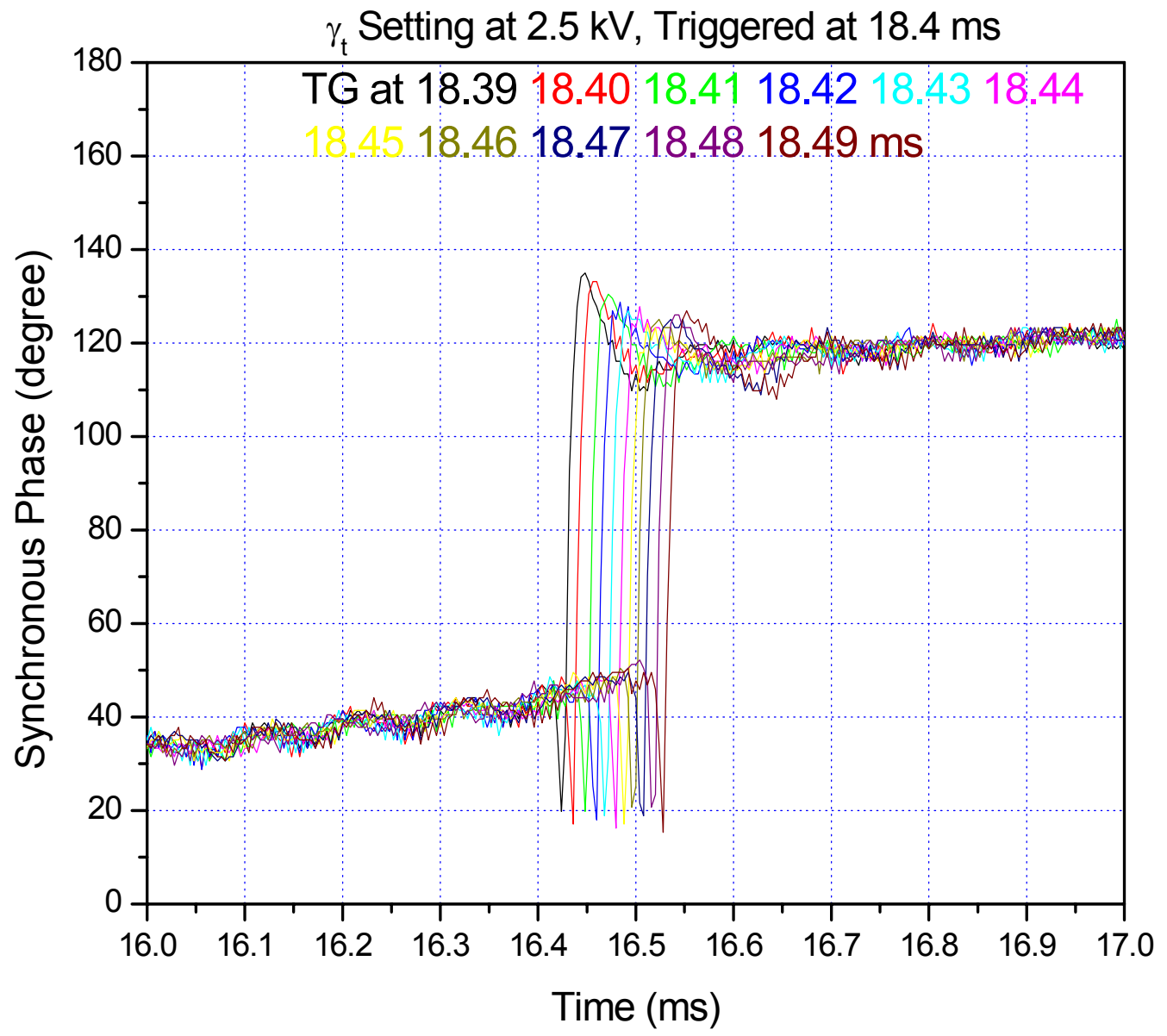

$\gamma_t$ Setting at 2.5 kV, Triggered at 18.4 ms
TG at 18.39 18.40 18.41 18.42 18.43 18.44
18.45 18.46 18.47 18.48 18.49 ms
Synchronous Phase (degree)
Time (ms)

Fig. 5(c)

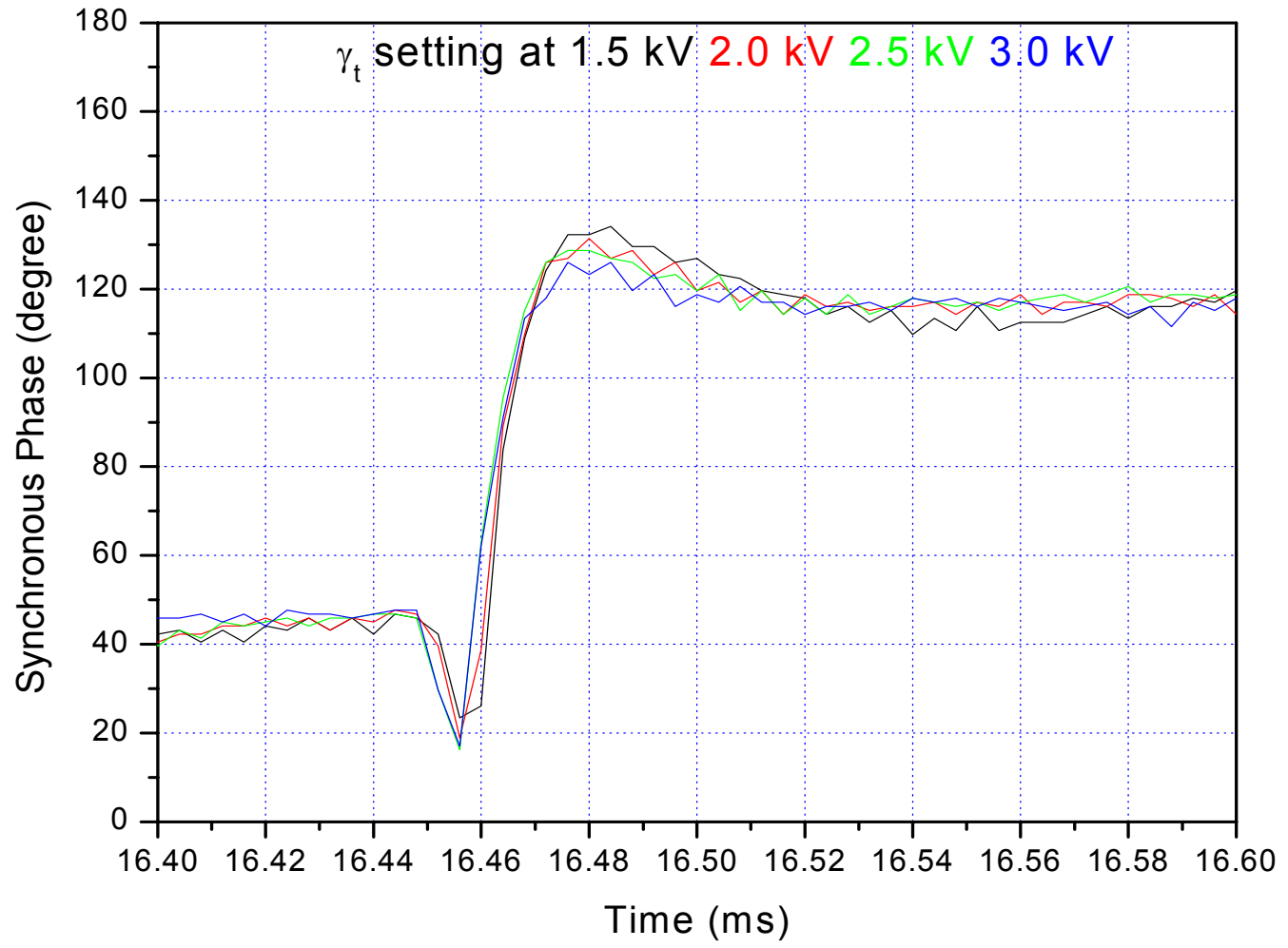

$\gamma_t$ setting at 1.5 kV 2.0 kV 2.5 kV 3.0 kV
Synchronous Phase (degree)
Time (ms)

Fig. 5(d)

Fig. 5(a) the optimal $T_G$ *vs*. $T_P$ at three different $V$ settings of 2.0 kV (the green triangle), 2.5 kV (the red circle), and 3.0 kV (the black square). The magenta curve represents $T_G=T_P$.

Fig. 5(b) the optimal delay of "$T_G$-$T_P$" *vs*. $V$ at the fixed $T_P$ of 18.4 ms.

Fig. 5(c) at $V$=2.5 kV and $T_P$=18.40 ms, the synchronous phase *vs*. time at eleven different $T_G$ of 18.39 ms, 18.4 ms, 18.41 ms, 18.42 ms, 18.43 ms, 18.44 ms, 18.45 ms, 18.46 ms, 18.47 ms, 18.48 ms, and 18.49 ms. The black, red, green, blue, cyan, magenta, yellow, dark yellow, navy, purple, wine curves represent the eleven different $T_G$ of 18.39 ms, 18.40 ms, 18.41 ms, 18.42 ms, 18.43 ms, 18.44 ms, 18.45 ms, 18.46 ms, 18.47 ms, 18.48 ms, and 18.49 ms respectively.

Fig. 5(d) at $T_P$=18.40 ms and $T_G$=18.42 ms, the synchronous phase *vs*. time at four different $V$ of 1.5 kV, 2.0 kV, 2.5 kV, and 3.0 kV. The black, red, green, and blue curves represent four different $V$ of 1.5 kV, 2.0 kV, 2.5 kV, and 3.0 kV respectively.

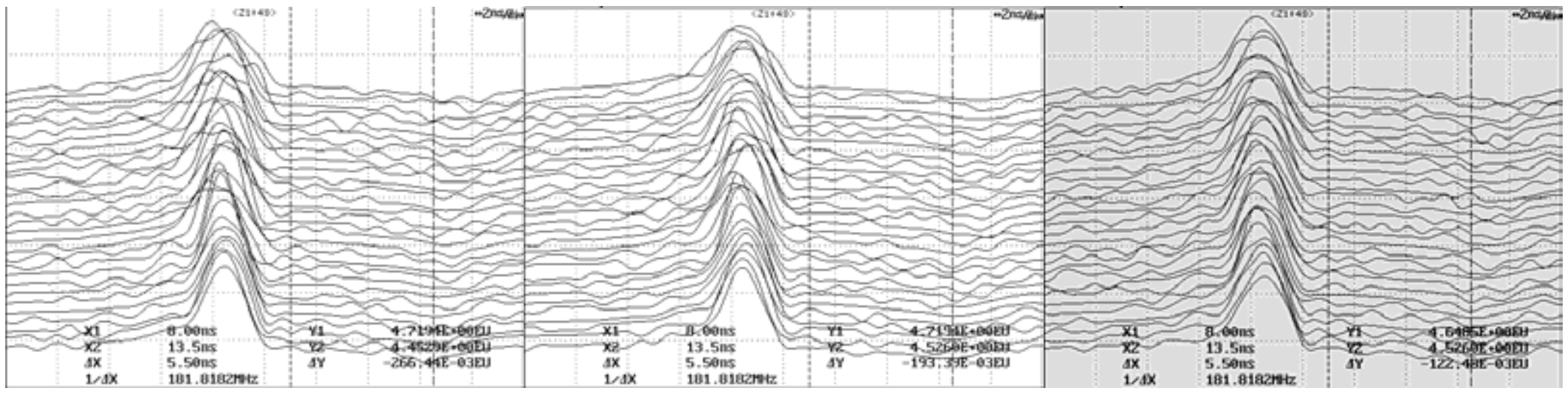

(a) (b) (c)

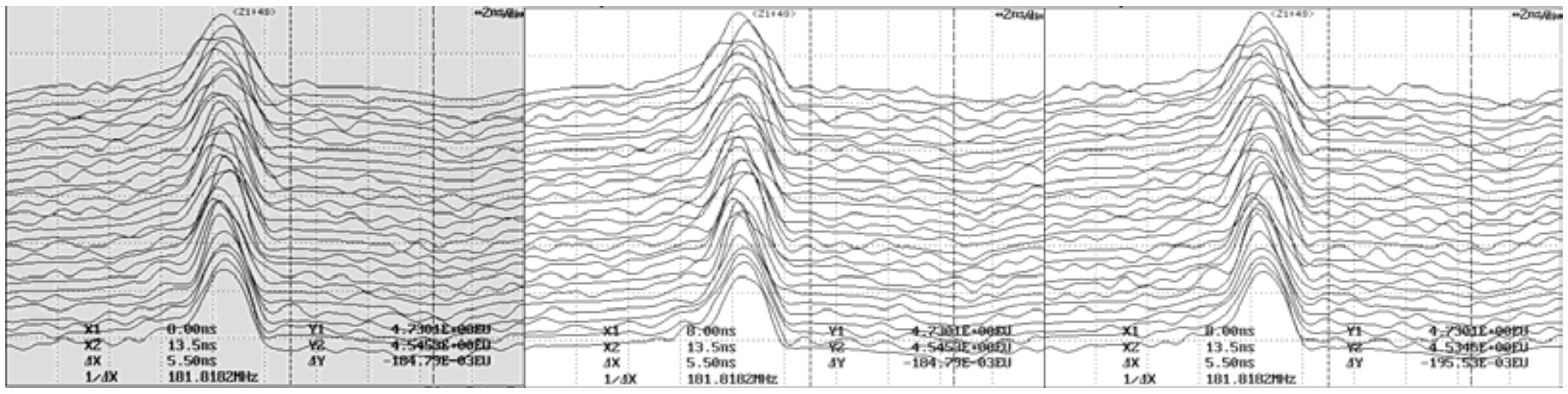

(d) (e) (f)

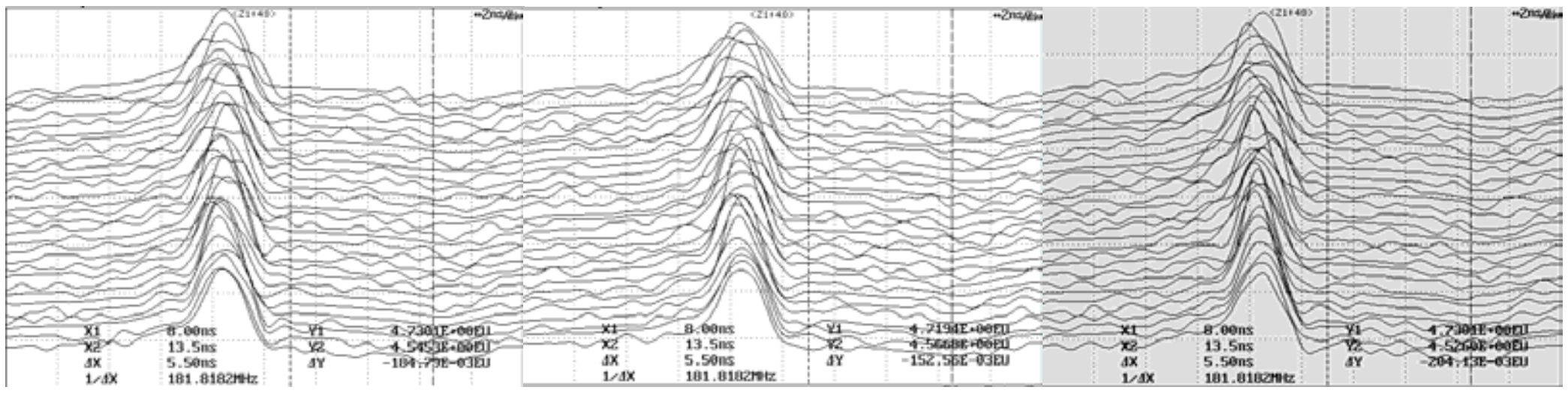

(g) (h) (i)

Fig. 6(a) at the extracted beam intensity of $4.0\times10^{12}$ protons, MRP took at the TJS setting of $V$=1.5 kV, $T_P$=18.70 ms and $T_G$=18.710 ms.

Fig. 6(b) MRP at the TJS setting of $V$=1.5 kV, $T_P$=18.70 ms and $T_G$=18.720 ms.

Fig. 6(c) MRP at the TJS setting of $V$=1.5 kV, $T_P$=18.70 ms and $T_G$=18.725 ms.

Fig. 6(d) MRP at the TJS setting of $V$=1.5 kV, $T_P$=18.70 ms and $T_G$=18.730 ms.

Fig. 6(e) MRP at the TJS setting of $V$=1.5 kV, $T_P$=18.70 ms and $T_G$=18.735 ms.

Fig. 6(f) MRP at the TJS setting of $V$=1.5 kV, $T_P$=18.70 ms and $T_G$=18.740 ms.

Fig. 6(g) MRP at the TJS setting of $V$=1.5 kV, $T_P$=18.70 ms and $T_G$=18.750 ms.

Fig. 6(h) MRP at the TJS setting of $V$=1.5 kV, $T_P$=18.70 ms and $T_G$=18.760 ms.

Fig. 6(i) MRP at the operational condition with $V$=0 kV, a different ROF, and transition gate $t_g$=18.980 ms.